\documentclass[conference]{IEEEtran}
\input epsf
\newcommand{\subvt}[2]{{#1}_{\text{#2}}}
\newcommand{\mtopgfa}[1]{\immediate\write18{m4.exe -dp -I "C:\string\Coding\string\Coding\string\LaTeX\string\circuit_macros" pgf.m4 #1.m4 | dpic -g > #1.pgf}}

\usepackage{graphicx}
\usepackage{xcolor}
\usepackage{amsmath}
\usepackage{amsfonts}
\usepackage{amssymb}
\usepackage{siunitx}
\usepackage{tikz}
\usepackage{pgfplots}

\begin{document}

%+++++++++++++++++++++++++++++++++++++++++++
\title{\LARGE Time-domain Representation of Passband Scattering Parameters}
%+++++++++++++++++++++++++++++++++++++++++++

\author{\authorblockN{{Justin B. King, \emph{Senior Member, IEEE}.}}

\authorblockA{Department of Electronic and Electrical Engineering, Trinity College Dublin, Ireland.}}

\newcommand{\cbl}{equivalent baseband}
\newcommand{\cb}{equivalent baseband}

%+++++++++++++++++++++++++++++++++++++++++++++++++++

\maketitle

\begin{abstract}
This paper presents a simple and accurate method for the inclusion of linear, time-invariant (LTI) networks, described by RF frequency-domain data, within \cbl{} time-domain simulations. The time-domain representation is formulated as an \cbl{} discrete-time impulse response, which may be convolved with the \cbl{} form of the input signal, to obtain the corresponding \cbl{} output. This allows networks which are most accurately described in the frequency domain, such as frequency-dispersive transmission lines, to be efficiently included as part of a transient time-domain simulation.
\end{abstract}
\IEEEoverridecommandlockouts
\begin{keywords}
Transient~simulation, convolution, S-parameters, behavioural modelling.
\end{keywords}
\IEEEpeerreviewmaketitle
\section{Introduction}
Simulation of nonlinear high-frequency circuits has traditionally focused on both pure transient and harmonic balance (HB) simulation techniques. While HB admits simple inclusion of frequency-domain descriptions, it is limited to steady-state solutions involving only quasiperidic signals. Transient simulation, on the other hand, is well-suited to dealing with highly nonlinear systems subject to complex digitally modulated excitations, with SPICE the archetype example of the power of this technique. For RF systems, however, it is not immediately clear how frequency-domain data (e.g. $S$-parameter descriptions) can be accommodated within such time-domain simulations.

Several techniques have been proposed to address this problem by carefully transforming baseband frequency-domain descriptions into a form that may be included within a transient simulation \cite{BrazilIMS2007, Gustavsen_VFIT_Complex}. In addition, previous work has shown that it is possible to extend \cite{Gustavsen_VFIT_Complex} to include RF passband descriptions within \cbl{} behavioural simulation frameworks \cite{KingBrazilINMMIC2017}. 

This paper presents an extension of \cite{Condon_FSE}, allowing tabulated passband frequency-domain data to be transformed to a complex-valued \cbl{}, discrete-time, impulse response which may be included within a transient solver.

\section{Frequency-domain representation}
% no \PARstart
Consider a passband scattering parameter description of an LTI one-port system $F(\omega)$, with bandwidth $2\subvt{\omega}{m}$. The \cbl{} form of this signal is defined as $\tilde{F}(\omega)$, and we note that it does not posses conjugate symmetry in general. At this point we may approximate $\tilde{F}(\omega)$ as the following frequency-domain complex Fourier series
%Note that for an $N$-port network, $F(\omega)$ is an $N \times N$ matrix.
\begin{IEEEeqnarray}{rCl}
	\label{eq:complex_fourier}
	\tilde{S}(\omega) = \sum_{k=0}^{\infty} s_{k} e^{-jk\Omega\omega},
\end{IEEEeqnarray}
where the $s_k$ are the Fourier coefficients, the frequency repetition `period' is $2\subvt{\omega}{m}$ and hence $\Omega = \pi/\subvt{\omega}{m}$, while the lower summation limit is zero, to ensure time-domain causality.
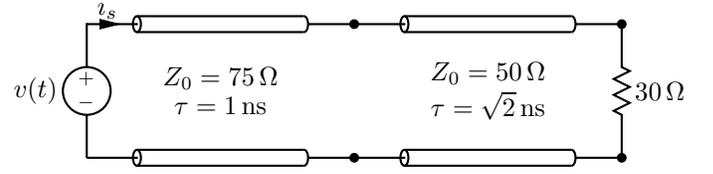
\begin{figure}
	\mtopgfa{tikz/transmission_line}%
	% lsm.m4

\begin{tikzpicture}[scale=2.54]
% dpic version 2017.12.01 option -g for TikZ and PGF 1.01
\ifx\dpiclw\undefined\newdimen\dpiclw\fi
\global\def\dpicdraw{\draw[line width=\dpiclw]}
\global\def\dpicstop{;}
\dpiclw=0.8bp
\dpiclw=0.8bp
\dpiclw=1bp
\dpiclw=0.8bp
\dpicdraw (0,0)
 --(0,0.225)\dpicstop
\dpicdraw (0,0.35) circle (0.049213in)\dpicstop
\draw (0,0.2875) node{$_-$};
\draw (0,0.4125) node{$_+$};
\dpicdraw (0,0.475)
 --(0,0.7)\dpicstop
\draw (-0.125,0.35) node[left=-1.5bp]{$ v(t)$};
\dpicdraw (0,0.694444)
 --(0,0.705556)\dpicstop
\dpicdraw (1.158333,0.7)
 --(1.4,0.7)\dpicstop
\dpicdraw (0,0.7)
 --(0.2625,0.7)\dpicstop
\dpicdraw[line width=0.4bp](0.2625,0.7) circle (0.00109in)\dpicstop
\dpicdraw (0.2625,0.658333)
 --(1.1375,0.658333)\dpicstop
\dpicdraw (1.1375,0.658333)
 ..controls (1.149006,0.658333) and (1.158333,0.676988)
 ..(1.158333,0.7)
 ..controls (1.158333,0.723013) and (1.149006,0.741667)
 ..(1.1375,0.741667)\dpicstop
\dpicdraw (1.1375,0.741667)
 --(0.2625,0.741667)\dpicstop
\dpicdraw (0.2625,0.741667)
 ..controls (0.250994,0.741667) and (0.241667,0.723013)
 ..(0.241667,0.7)
 ..controls (0.241667,0.676988) and (0.250994,0.658333)
 ..(0.2625,0.658333)
 ..controls (0.274006,0.658333) and (0.283333,0.676988)
 ..(0.283333,0.7)
 ..controls (0.283333,0.723013) and (0.274006,0.741667)
 ..(0.2625,0.741667)\dpicstop
\dpicdraw[fill=black](1.4,0.7) circle (0.007874in)\dpicstop
\filldraw[line width=0bp](0.069333,0.675)
 --(0.169333,0.7)
 --(0.069333,0.725) --cycle\dpicstop
\dpicdraw (0.146427,0.7)
 --(0.069333,0.7)\dpicstop
\draw (0.10788,0.7) node[above=-1.5bp]{$ i_s$};
\dpicdraw (1.158333,0)
 --(1.4,0)\dpicstop
\dpicdraw (0,0)
 --(0.2625,0)\dpicstop
\dpicdraw[line width=0.4bp](0.2625,0) circle (0.00109in)\dpicstop
\dpicdraw (0.2625,-0.041667)
 --(1.1375,-0.041667)\dpicstop
\dpicdraw (1.1375,-0.041667)
 ..controls (1.149006,-0.041667) and (1.158333,-0.023013)
 ..(1.158333,0)
 ..controls (1.158333,0.023013) and (1.149006,0.041667)
 ..(1.1375,0.041667)\dpicstop
\dpicdraw (1.1375,0.041667)
 --(0.2625,0.041667)\dpicstop
\dpicdraw (0.2625,0.041667)
 ..controls (0.250994,0.041667) and (0.241667,0.023013)
 ..(0.241667,0)
 ..controls (0.241667,-0.023013) and (0.250994,-0.041667)
 ..(0.2625,-0.041667)
 ..controls (0.274006,-0.041667) and (0.283333,-0.023013)
 ..(0.283333,0)
 ..controls (0.283333,0.023013) and (0.274006,0.041667)
 ..(0.2625,0.041667)\dpicstop
\dpicdraw[fill=black](1.4,0) circle (0.007874in)\dpicstop
\draw (0.7,0.35) node{\shortstack{$Z_{0} = \SI{75}{\ohm}$\\%
$\tau = \SI{1}{\nano\second}$}};
\dpicdraw (2.558333,0.7)
 --(2.8,0.7)\dpicstop
\dpicdraw (1.4,0.7)
 --(1.6625,0.7)\dpicstop
\dpicdraw[line width=0.4bp](1.6625,0.7) circle (0.00109in)\dpicstop
\dpicdraw (1.6625,0.658333)
 --(2.5375,0.658333)\dpicstop
\dpicdraw (2.5375,0.658333)
 ..controls (2.549006,0.658333) and (2.558333,0.676988)
 ..(2.558333,0.7)
 ..controls (2.558333,0.723013) and (2.549006,0.741667)
 ..(2.5375,0.741667)\dpicstop
\dpicdraw (2.5375,0.741667)
 --(1.6625,0.741667)\dpicstop
\dpicdraw (1.6625,0.741667)
 ..controls (1.650994,0.741667) and (1.641667,0.723013)
 ..(1.641667,0.7)
 ..controls (1.641667,0.676988) and (1.650994,0.658333)
 ..(1.6625,0.658333)
 ..controls (1.674006,0.658333) and (1.683333,0.676988)
 ..(1.683333,0.7)
 ..controls (1.683333,0.723013) and (1.674006,0.741667)
 ..(1.6625,0.741667)\dpicstop
\dpicdraw[fill=black](2.8,0.7) circle (0.007874in)\dpicstop
\dpicdraw (2.558333,0)
 --(2.8,0)\dpicstop
\dpicdraw (1.4,0)
 --(1.6625,0)\dpicstop
\dpicdraw[line width=0.4bp](1.6625,0) circle (0.00109in)\dpicstop
\dpicdraw (1.6625,-0.041667)
 --(2.5375,-0.041667)\dpicstop
\dpicdraw (2.5375,-0.041667)
 ..controls (2.549006,-0.041667) and (2.558333,-0.023013)
 ..(2.558333,0)
 ..controls (2.558333,0.023013) and (2.549006,0.041667)
 ..(2.5375,0.041667)\dpicstop
\dpicdraw (2.5375,0.041667)
 --(1.6625,0.041667)\dpicstop
\dpicdraw (1.6625,0.041667)
 ..controls (1.650994,0.041667) and (1.641667,0.023013)
 ..(1.641667,0)
 ..controls (1.641667,-0.023013) and (1.650994,-0.041667)
 ..(1.6625,-0.041667)
 ..controls (1.674006,-0.041667) and (1.683333,-0.023013)
 ..(1.683333,0)
 ..controls (1.683333,0.023013) and (1.674006,0.041667)
 ..(1.6625,0.041667)\dpicstop
\dpicdraw[fill=black](2.8,0) circle (0.007874in)\dpicstop
\draw (2.1,0.35) node{\shortstack{$Z_{0} = \SI{50}{\ohm}$\\%
$\tau = \sqrt{2}\,\si{\nano\second}$}};
\dpicdraw (2.8,0.7)
 --(2.8,0.475)
 --(2.841667,0.454167)
 --(2.758333,0.4125)
 --(2.841667,0.370833)
 --(2.758333,0.329167)
 --(2.841667,0.2875)
 --(2.758333,0.245833)
 --(2.8,0.225)
 --(2.8,0)\dpicstop
\draw (2.841667,0.35) node[right=-1.5bp]{$ \SI{30}{\ohm}$};
\end{tikzpicture}%
	\caption{One-port network which is to be represented in the frequency domain (see Fig.~\ref{fig:freq_approx} for frequency response). Note how the delay of the second transmission line has been chosen such that the total response is not periodic.}
	\label{fig:tline}%
\end{figure}
The frequency-domain \cbl{} scattered response $\tilde{B}(\omega)$ to an incident pseudowave $\tilde{A}(\omega)$, is given as
\begin{IEEEeqnarray}{rCl}
	\label{eq:pseudowave_transfer_function}
	\tilde{B}(\omega) = \tilde{S}(\omega)\tilde{A}(\omega).
\end{IEEEeqnarray}
The time-domain impulse response may now be obtained through an inverse Fourier transform
\begin{IEEEeqnarray}{rCl}
	\label{eq:pseudowave_transfer_function}
	\tilde{h}(t) &=& \frac{1}{2\pi}\int_{-\infty}^{\infty} \tilde{S}(\omega)\cdot 1 \cdot  e^{j\omega t} \, \mathrm{d}\omega \IEEEnonumber \\
	&=&  \sum_{k=0}^{\infty} s_{k} \frac{1}{2\pi}\int_{-\infty}^{\infty}  e^{j\omega(t - \tilde{k})} \, \mathrm{d}\omega \\
	&=&  \sum_{k=0}^{\infty} s_{k} \delta(t-\tilde{k}), \IEEEnonumber
\end{IEEEeqnarray}
where $\tilde{k} = k\pi/\subvt{\omega}{m}$.
%\tilde{k}

The expression for $\tilde h(t)$ above may be interpreted as a discrete-time impulse response $\tilde{s}[k]=\tilde{h}(k\pi/\subvt{\omega}{m})$, with uniform time-step of $\pi/\subvt{\omega}{m}$. Note that the impulse response weights $s_{k}$ are complex, since this represents the \cbl{} form of the original passband network. In general for a multiport network, $\tilde{s}[k]$ could be a $P \times P$ matrix of discrete-time impulse response sequences. It is noted that, in practice, the Fourier series expansion in (\ref{eq:complex_fourier}) must be limited to some finite upper harmonic index $N$.

The impulse response $\tilde{s}[k]$ may now be included as part of a transient simulation, via a discrete-time complex convolution operation
\begin{IEEEeqnarray}{rCl}
	\label{eq:complex_convolution}
	\tilde{b}[n] = (\tilde{s} \ast \tilde{a})[n] = \tilde{s}[0]\tilde{a}[n] + \sum_{k = 1}^{n} \tilde{s}[k]\tilde{a}[n-k].
\end{IEEEeqnarray}
The terms within the summation symbol above represent the past convolution history, and thus contain only known data, while the first term contains the known impulse response $\tilde{s}[0]$, along with the unknown incident pseudowave $\tilde{a}[n]$, which must be solved for at each new time-step.

\section{Extraction of Fourier coefficients}
Consider tabulated \cbl{} \emph{S}-parameter data $\tilde{F}(\omega_i)$ with compact support in the range $[-\subvt{\omega}{m},\subvt{\omega}{m}]$, written as the column vector,
\begin{IEEEeqnarray}{rCl}
	\label{eq:rhs_F_vector}
	\mathbf{F} = 
	\begin{bmatrix}
		\tilde{F}(-\subvt{\omega}{m})\\
		\vdots\\
		\tilde{F}(\omega_i)\\
	    \vdots\\
		\tilde{F}(\subvt{\omega}{m})\\
	\end{bmatrix},
\end{IEEEeqnarray}
and let us also define the following coefficient matrix
\begin{IEEEeqnarray}{rCl}
	\label{eq:M_matrix}
	\mathbf{M} = 
	\begin{bmatrix}
		1 & \cdots & \exp(j\Omega k \subvt{\omega}{m}) & \cdots & \exp(j\Omega N \subvt{\omega}{m})\\
		\vdots & \ddots & \vdots &  \\
		1 & \cdots & \exp(j\Omega k \omega_i) & \cdots & \exp(j\Omega N \omega_i)\\
		\vdots & & \vdots &  \ddots\\
		1 & \cdots & \exp(-j\Omega k \subvt{\omega}{m}) & \cdots & \exp(-j\Omega N \subvt{\omega}{m})\\
	\end{bmatrix}. \IEEEeqnarraynumspace
\end{IEEEeqnarray}
Equation (\ref{eq:complex_fourier}) may now be expressed as
\begin{IEEEeqnarray}{rCl}
	\label{eq:least_squares_problem}
	\mathbf{M} \mathbf{s} + \mathbf{E} = \mathbf{F},
\end{IEEEeqnarray}
where $\mathbf{s}$ is the column vector of impulse response coefficients, $s_k$, and $\mathbf{E}$ is the Fourier approximation error. This error is minimised in a least squares sense by taking
\begin{IEEEeqnarray}{rCl}
	\label{eq:least_squares_solution}
	\mathbf{s} = (\mathbf{M}^\mathsf{H} \cdot \mathbf{M})^{-1} \cdot \mathbf{M}^\mathsf{H} \cdot \mathbf{F},
\end{IEEEeqnarray}
i.e. the weights $s_{k}$ can be extracted through a simple matrix operation. Note that $\mathbf{M}^{\mathsf{H}}$ represents the conjugate transpose of $\mathbf{M}$. In the next section we validate the proposed method through a simple practical example.
\section{Practical example}
\begin{figure}
	\begin{tikzpicture}

%\path[use as bounding box] (-1, -1.2) rectangle (7.9, 3.75);

\pgfplotsset{
        table/search path={tikz},
}

\begin{axis}[
name=real,
scale only axis,
xmin=0, xmax=20,
xtick = {0,2,...,20},
ymin = -0.2, ymax = 0.3,
ytick = {-0.2,-0.1,...,0.3},
yticklabels = {-0.2,,0,,,0.3},
%yticklabels = {-0.25,,,,,0.25,},
%extra y ticks={0},
%extra y tick labels={0},
xlabel near ticks,
ylabel near ticks,
width = 7cm,
height = 2.5cm,
xlabel={Sample ($k$)},
ylabel={$\tilde{s}[k]$ (unitless)},
xmajorgrids,
ymajorgrids,
xminorgrids,
yminorgrids,
legend cell align=left,
clip=false,
restrict x to domain=0:20, 
]
\addplot [ycomb, color=red, line width=1.0pt, mark = square*, mark size=1] plot table[x expr=\coordindex, y expr=\thisrowno{0}] {cir.txt};
\addplot [ycomb, color=blue, dash pattern=on 1pt off 1pt, line width=1.0pt, mark = square*,mark options={solid}, mark size=1] plot table[x expr=\coordindex, y expr=\thisrowno{1}] {cir.txt};
\legend{{Re\,$s_k$}, {Im\,$s_k$}}
\end{axis}

\end{tikzpicture}%
	\caption{The real and imaginary parts of the lowpass equivalent discrete-time impulse response. The rapid decay of the response is clearly evident.}
	\label{fig:dtir}%
\end{figure}
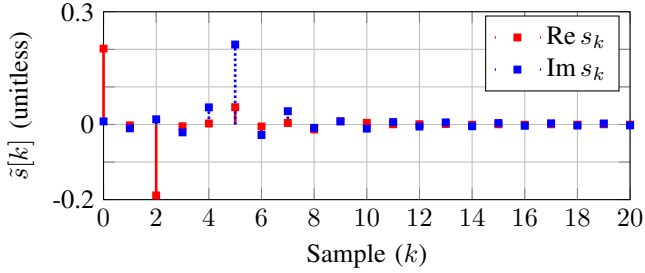

\begin{figure}
	\begin{tikzpicture}

%\path[use as bounding box] (-1, -1.2) rectangle (7.9, 3.75);
\pgfplotsset{
        table/search path={tikz},
}
\begin{axis}[
name=real,
scale only axis,
xmin=-0.5, xmax=0.5,
ymin = -0.5, ymax = 0.7,
ytick = {-0.5, -0.3,-0.1,0.1, 0.3,0.5,0.7},
yticklabels = {-0.5,,0,,0.3,,0.6},
xlabel near ticks,
ylabel near ticks,
width = 7cm,
height = 2.5cm,
xlabel={Frequency (GHz)},
ylabel={Amplitude},
xmajorgrids,
ymajorgrids,
xminorgrids,
yminorgrids,
legend cell align=left,
legend pos = south east,
mark repeat={8},
]
\addplot [color=red, line width=0.75pt] plot table[x index=0, y index=1] {freq_approx.txt}; 
\addplot [line width=0.75pt, only marks, black, mark = asterisk] plot table[x index=0, y index=2] {freq_approx.txt};
\addplot [color=blue, line width=0.75pt] plot table[x index=0, y index=3] {freq_approx.txt}; 
\addplot [line width=0.75pt, only marks, green!40!black, mark = x] plot table[x index=0, y index=4] {freq_approx.txt};
%\legend{$\tilde{S}(\omega)$, {Im\,$s_k$}}
\legend{Re\,$\tilde{F}(\omega)$,Re\,$\tilde{S}(\omega)$,Im\,$\tilde{F}(\omega)$,Im\,$\tilde{S}(\omega)$}
\end{axis}

\end{tikzpicture}
	\caption{The real and imaginary parts of the downshifted frequency-domain data (red and blue lines) compared with the frequency-domain Fourier series approximation (black and green symbols).}
	\label{fig:freq_approx}
\end{figure}
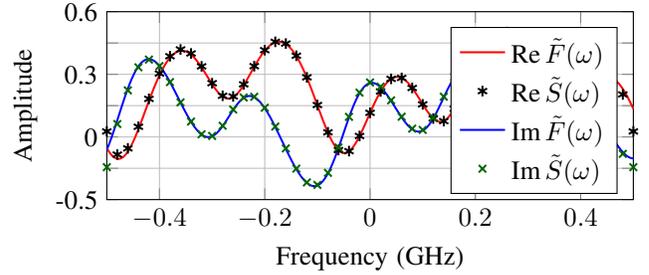

Consider the non-commensurate transmission line network shown in Fig.~\ref{fig:tline}.
The $S$-parameters of this one-port have been measured over a frequency range from $\SI{9.5}{\giga\hertz}$ to $\SI{10.5}{\giga\hertz}$, and a complex Fourier expansion in the form (\ref{eq:complex_fourier}) has been extracted via (\ref{eq:least_squares_solution}) to yield a complex valued discrete-time impulse response $\tilde{s}[k]$, which is shown in Fig.~\ref{fig:dtir}.
The frequency-domain approximation this admits is shown in Fig.~\ref{fig:freq_approx}. 

\begin{figure}
	\begin{tikzpicture}

%\path[use as bounding box] (-1, -1.2) rectangle (7.9, 3.75);
\pgfplotsset{
        table/search path={tikz},
}

\begin{axis}[
name=real,
scale only axis,
xmin=0, xmax=50,
ymin=-40, ymax=60,
xlabel near ticks,
ylabel near ticks,
width = 7cm,
height = 2.5cm,
xlabel={Time (ns)},  
ylabel={$i_s$ (mA)},
xmajorgrids,
ymajorgrids,
xminorgrids,
yminorgrids,
legend cell align=left,
legend pos = north east,
%mark repeat={4}, 
restrict x to domain=0:50,
]
\addplot [color=blue, line width=0.75pt, samples = 1000] plot table[x index=0, y index=1] {output_ads.txt}; 
\addplot [ycomb,  line width=0.75pt, color=red, mark = x] plot table[x index=0, y index=1] {output_conv.txt};
%\legend{$\tilde{S}(\omega)$, {Im\,$s_k$}} 
\legend{Commercial Solver, This Paper} 
\end{axis}

\end{tikzpicture}
	\caption{Comparison of the downconverted time-domain output current waveform $i_s$ (see Fig.~\ref{fig:tline}) from the commercial solver (blue line) versus the baseband time domain output from the technique described in this paper (red lines, crosses).}
	\label{fig:output_time}
\end{figure}
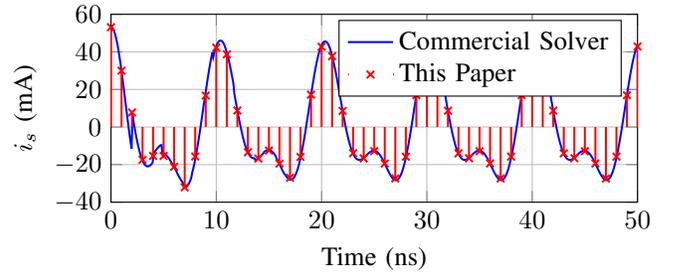

The transmission line network of Fig.~\ref{fig:tline} is now excited with a multisine input (four equal amplitude tones, equally spaced between $\SI{9.8}{\giga\hertz}$ and $\SI{10.2}{\giga\hertz}$). The current $i_s$ into the line is now solved for in the transient regime using a commercial simulation environment, and an in-house transient solver implementing the \cbl{} convolution method of this paper. Note that the commercial solver operates at passband frequencies, and requires a built-in model for the  transmission line. The baseband time-domain output of both solvers is shown in Fig.~\ref{fig:output_time}, with excellent agreement visible.
\section{Conclusion}
A method for the inclusion of passband frequency-domain data within transient simulators is presented. In general, the time-domain description is very compact (Fig.~\ref{fig:dtir}). The time domain solution shows excellent agreement compared to the results from a commercial solver operating at much faster time-step (Fig.~\ref{fig:output_time}).

\bibliographystyle{IEEEtran}
\bibliography{mypapers,references,jkbooks,jkothers,jkpapers}

\begin{thebibliography}{1}
\providecommand{\url}[1]{#1}
\csname url@samestyle\endcsname
\providecommand{\newblock}{\relax}
\providecommand{\bibinfo}[2]{#2}
\providecommand{\BIBentrySTDinterwordspacing}{\spaceskip=0pt\relax}
\providecommand{\BIBentryALTinterwordstretchfactor}{4}
\providecommand{\BIBentryALTinterwordspacing}{\spaceskip=\fontdimen2\font plus
\BIBentryALTinterwordstretchfactor\fontdimen3\font minus
  \fontdimen4\font\relax}
\providecommand{\BIBforeignlanguage}[2]{{%
\expandafter\ifx\csname l@#1\endcsname\relax
\typeout{** WARNING: IEEEtran.bst: No hyphenation pattern has been}%
\typeout{** loaded for the language `#1'. Using the pattern for}%
\typeout{** the default language instead.}%
\else
\language=\csname l@#1\endcsname
\fi
#2}}
\providecommand{\BIBdecl}{\relax}
\BIBdecl

\bibitem{BrazilIMS2007}
T.~J. Brazil, ``Nonlinear, transient simulation of distributed {RF} circuits
  using discrete-time convolution,'' in \emph{2007}, Jun. 2007, pp. 505--508.

\bibitem{Gustavsen_VFIT_Complex}
B.~Gustavsen and A.~Semlyen, ``Rational approximation of frequency domain
  responses by vector fitting,'' \emph{IEEE Transactions on Power Delivery},
  vol.~14, no.~3, pp. 1052--1061, Jul 1999.

\bibitem{KingBrazilINMMIC2017}
J.~B. King and T.~J. Brazil, ``Time-domain simulation of passband {S}-parameter
  networks using complex baseband vector fitting,'' in \emph{2017 Integrated
  Nonlinear Microwave and Millimetre-wave Circuits, Graz, Austria.}, Apr. 2017.

\bibitem{Condon_FSE}
M.~Condon, R.~Ivanov, and C.~Brennan, ``A causal model for linear {RF} systems
  developed from frequency-domain measured data,'' \emph{IEEE Trans. Circuits
  \& Systems II: Express Briefs}, vol.~52, no.~8, pp. 457--460, Aug. 2005.

\end{thebibliography}
\end{document}